\documentclass{elsart}

\usepackage{cite}
\usepackage{graphicx}

\begin{document}

\begin{frontmatter}

\title{On the small--amplitude approximation to the differential equation
$\ddot{x}+(1+\dot{x}^{2})x=0$}
\author{Francisco M. Fern\'{a}ndez \thanksref{FMF}}

\address{INIFTA (UNLP, CCT La Plata-CONICET), Divisi\'{o}n Qu\'{i}mica Te\'{o}rica,\\
Diag. 113 y 64 (S/N), Sucursal 4, Casilla de Correo 16,\\
1900 La Plata, Argentina}

\thanks[FMF]{e--mail: fernande@quimica.unlp.edu.ar}

\begin{abstract}
We obtain the radius of convergence of the small--amplitude approximation to
the period of the nonlinear oscillator $\ddot{x}+(1+\dot{x}^{2})x=0$ with
the initial conditions $x(0)=A$ and $\dot{x}(0)=0$ and show that the
inverted perturbation series appears to converge smoothly from below.
\end{abstract}

\end{frontmatter}

There has recently been great interest in the study of the period of the
nonlinear oscillator
\begin{eqnarray}
&&\ddot{x}(t)+[1+\dot{x}(t)^{2}]x(t)=0  \nonumber \\
&&x(0)=A,\;\dot{x}(0)=0  \label{eq:difeq_x}
\end{eqnarray}
as a function of the amplitude $A$. Apparently, it aroused from the fact
that the first--order harmonic balance method yielded the approximate
frequency\cite{C03}
\begin{equation}
\omega ^{HB}(A)=\frac{2}{\sqrt{4-A^{2}}}  \label{eq:omega_HB}
\end{equation}
that is not defined for $A>2$.

By straightforward analysis of the dynamical trajectories in the $x-y$
plane, where $y=\dot{x}$, Beatty and Mickens\cite{BM05} concluded that such
a restriction is merely an artifact of the harmonic balance method.

Later, Mickens\cite{M06} derived an explicit expression for the period
\begin{equation}
T(A)=4A\int_{0}^{1}\frac{du}{\sqrt{e^{A^{2}(1-u^{2})}-1}}  \label{eq:T(A)}
\end{equation}
where $u=x/A$. By means of this expression he proved that $dT/dA<0$ and
obtained upper and lower bounds to the period.

Kalm\'{a}r--Nagy and Erneux\cite{KE08} derived the behaviour of the period
for small and large values of $A$%
\begin{eqnarray}
T(A) &\simeq &2\pi \left( 1-\frac{A^{2}}{8}\right) ,\;A\ll 1  \nonumber \\
T(A) &\simeq &\frac{2\pi }{A},\;A\gg 1  \label{eq:T(A)_asymp}
\end{eqnarray}
as well as most interesting approximations to the periodic orbits in both
limits. In particular, they showed that the trajectory $u(t)$ satisfies the
equation
\begin{eqnarray}
\ddot{u}+\frac{dV}{du} &=&0,  \nonumber \\
V(u) &=&\frac{1-e^{\rho (1-u^{2})}}{2\rho },\;\rho =A^{2}  \label{eq:V(u)}
\end{eqnarray}
that leads to the same expression for the period (\ref{eq:T(A)}) derived
earlier by Mickens\cite{M06}.

The results of those authors clearly show that the period $T(A)$ does not
exhibit singular points for real values of $A$ but they do not explain why
the harmonic balance fails as shown in equation (\ref{eq:omega_HB})\cite{C03}%
. A possible explanation is that the harmonic balance is reflecting a
singular point in the complex $A$--plane. If it exists, then the
small--amplitude expansion will have a finite radius of convergence.

In order to derive the small--amplitude expansion we change the integration
variable in equation (\ref{eq:T(A)}) to $u=\cos \theta $ so that the period
becomes
\begin{equation}
T(\rho )=4\int_{0}^{\pi /2}\frac{d\theta }{\sqrt{F(\rho \sin ^{2}\theta )}}
\label{eq:T(rho)}
\end{equation}
where
\begin{equation}
F(z)=\frac{e^{z}-1}{z}=\sum_{j=0}^{\infty }\frac{z^{j}}{(j+1)!}
\label{eq:F(z)}
\end{equation}
If we substitute the expansion
\begin{equation}
\frac{1}{\sqrt{F(z)}}=\sum_{j=0}^{\infty }c_{j}z^{j}=1-\frac{z}{4}+\frac{%
z^{2}}{96}+\frac{z^{3}}{384}+\ldots  \label{eq:1/sqrt(F)_series}
\end{equation}
into the equation (\ref{eq:T(rho)}) we obtain as many coefficients as
desired of the small--amplitude series
\begin{eqnarray}
T(A) &=&T_{0}+T_{1}\rho +T_{2}\rho ^{2}+\ldots  \nonumber \\
&=&2\pi \left( 1-\frac{\rho }{8}+\frac{\rho ^{2}}{256}+\ldots \right)
\label{eq:T_rho_series}
\end{eqnarray}

The function $1/\sqrt{e^{z}-1}$ has two complex--conjugate singular points
closest to the origin at $z=\pm 2\pi i$; therefore
\begin{equation}
\lim_{j\rightarrow \infty }\left| \frac{c_{j}}{c_{j+1}}\right| =2\pi
\end{equation}
If we take into account that
\begin{equation}
I_{j}=\int_{0}^{\pi /2}\sin ^{2j}\theta \,d\theta =\frac{\sqrt{\pi }\Gamma
(j+1/2)}{2\Gamma (j+1)}
\end{equation}
then we conclude that
\begin{equation}
\lim_{j\rightarrow \infty }\left| \frac{c_{j}\,I_{j}}{c_{j+1}\,I_{j+1}}%
\right| =\lim_{j\rightarrow \infty }\left| \frac{c_{j}}{c_{j+1}}\right| =2\pi
\end{equation}
In other words, the $\rho$--power series has a finite radius of convergence $%
R_{\rho }=2\pi $ because of a pair of complex conjugate singular points at $%
\rho _{c}=\pm 2\pi i$.

\begin{figure}[H]
\begin{center}
\includegraphics[width=9cm]{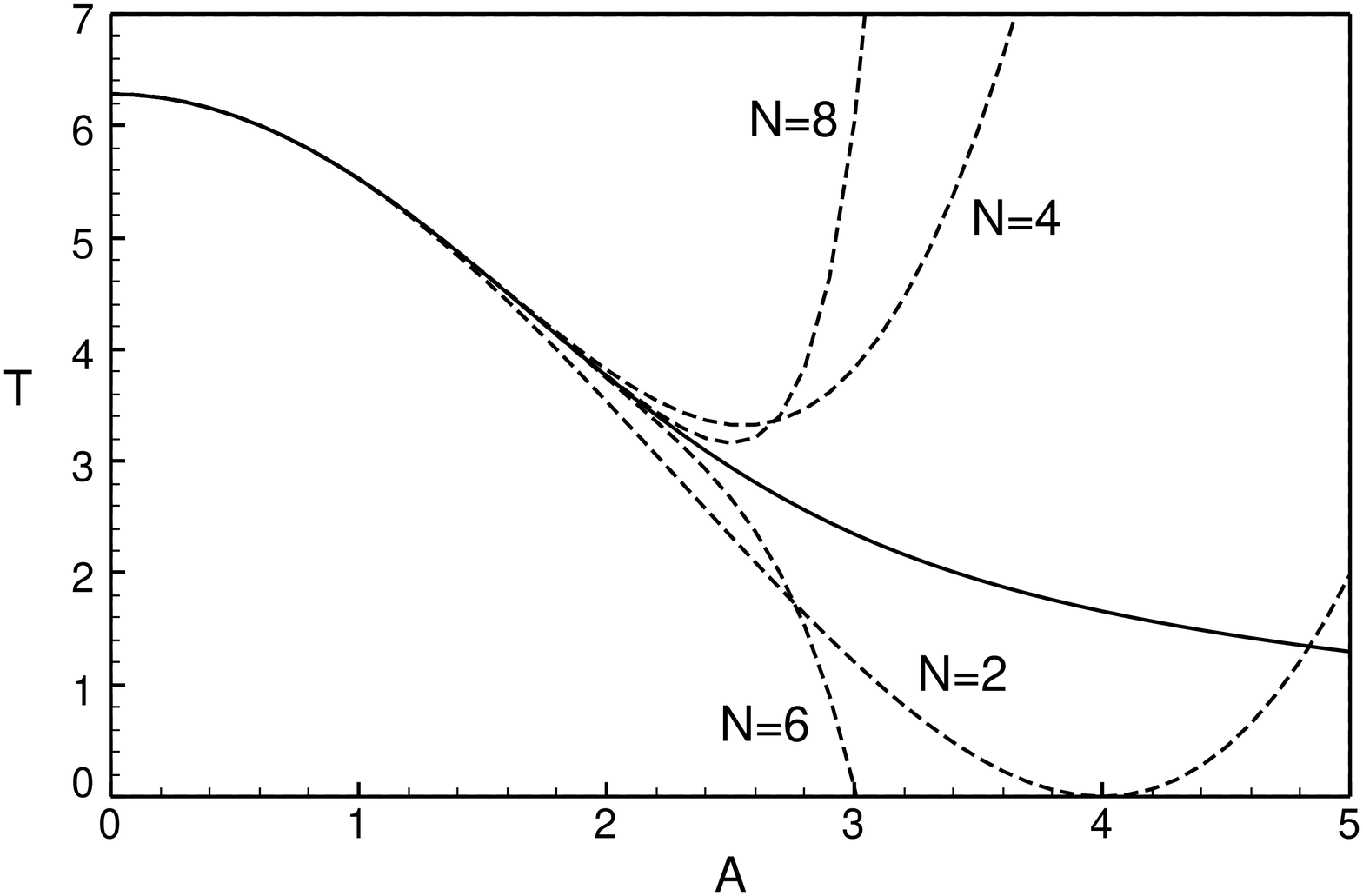}
\end{center}
\caption{Accurate numerical period (solid line) and partial sums for the
small--amplitude approximation (dashed lines).}
\label{fig:PT}
\end{figure}

Fig.~\ref{fig:PT} shows the first partial sums $S_{T}^{[N]}(\rho
)=T_{0}+T_{1}\rho +\ldots +T_{N}\rho ^{N}$ and the accurate numerical values
of $T(A)$. It dramatically illustrates the effect of the nonzero convergence
radius $R_{A}=\sqrt{2\pi }$ of the small--amplitude expansion determined by
the singular points of $T(A)$ closest to the origin in the complex $A$%
--plane.

We can provide another argument about the location of the singular points of
$T(A)$. First, note that the effective potential--energy function $V(u)$
given in equation (\ref{eq:V(u)}) exhibits a minimum $V(0)=(1-e^{\rho
})/(2\rho )<0$ and that the energy of the oscillatory motion is $E=\dot{u}%
^{2}/2+V(u)=0$ for the given initial conditions. Therefore, we expect a
critical value of $\rho $ given by $V(0)=0$ that yields $\rho _{c}=\pm 2\pi
i $ in agreement with the analysis above based on the small--amplitude
series.

In order to verify those exact analytical results in a numerical way we
constructed Pad\'{e} approximants $[N,N](\rho )$\cite{BO78} from the partial
sums $S_{T}^{[2N]}(\rho )$ and looked for the complex zeroes of the
denominator. A sequence of such zeroes appeared to converge to a limit quite
close to $\pm 6.3i$ with a small real part that was negligible compared to
the errors of the estimates. Besides, assuming that there is an algebraic
singular point\cite{BO78} closest to origin of the form $(z-z_{0})^{\alpha }$
we carried out the same Pad\'{e} analysis, but now on $T^{-1}dT/dA$ (as a
function of $\rho$), and obtained roughly the same complex numbers that are
quite close to $\pm 2\pi i$. Therefore, there appears to be no doubt that
the radius of convergence of the $\rho $--power series is in fact $R_{\rho
}=2\pi $ and is due to complex conjugate singular points located on the
imaginary axis of the complex $\rho $--plane at $\pm 2\pi i$.

\begin{figure}[H]
\begin{center}
\includegraphics[width=9cm]{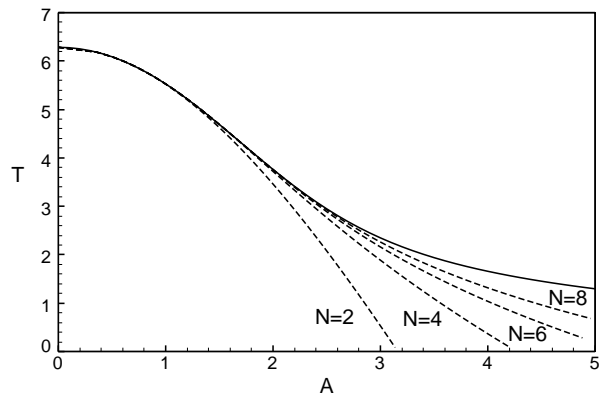}
\end{center}
\caption{Accurate numerical period (solid line) and partial sums for the
inverse perturbation series (dashed lines).}
\label{fig:IPT}
\end{figure}

Recently, Amore and Fern\'{a}ndez\cite{AF08} investigated the possible
advantages of the inverted perturbation series that in the present case
takes the form
\begin{eqnarray}
\rho &=&\rho _{1}\Delta T+\rho _{2}\Delta T^{2}+\ldots  \nonumber \\
&=&-\frac{4\Delta T}{\pi }+\frac{\Delta T^{2}}{2\pi ^{2}}-\frac{13\Delta
T^{3}}{24\pi ^{3}}+\ldots  \nonumber \\
\Delta T &=&T-2\pi  \label{eq:rho_DT_series}
\end{eqnarray}
Fig.~\ref{fig:IPT} shows that the partial sums for the inverted series $%
S_{\rho }^{[N]}(\Delta T)=\rho _{1}\Delta T+\rho _{2}\Delta T^{2}+\ldots
+\rho _{N}\Delta T^{N}$ converge smoothly from below towards the accurate
numerical values of $T(A)$. We are presently unable to prove such most
interesting feature of the inverted series rigorously.

\textbf{Summarizing}: Earlier studies on the nonlinear oscillator (\ref{eq:difeq_x})\cite
{BM05,M06,KE08} have clearly shown that the period is finite for all values
of the amplitude. However, they did not cast any light on the failure of the
harmonic balance (with the ansatz $x^{HB}(t)=A\cos (\omega t)$) that
predicts a singularity for $A=2$. In this paper we suggest that the harmonic
balance may be reflecting the singular points that determine the radius of
convergence of the small--amplitude series for the period. We have exactly
calculated the location of those singular points and concluded that the
series converge for $0<A<\sqrt{2\pi }$. This result may help to understand
similar difficulties in future applications of the harmonic balance. We
expect that a harmonic--balance approach with more terms will give a
frecuency with only complex singular points.

In addition to what was mentioned above, we have shown that for this problem
the inverted perturbation series appears to converge smoothly from below and
it is therefore preferable to the original small--amplitude expansion. It is
an old and well--known approach that may, in some cases, lead to
surprisingly accurate results\cite{AF08}.

\end{document}